\documentclass[onecolumn,preprintnumbers,amsmath,amssymb]{revtex4}
\usepackage{graphicx,caption,subcaption,float}
\begin{document}

\title{From Social to Epidemic Criticality and Back}

\author{Paolo Grigolini, David Lambert, Korosh Mahmoodi, Nicola Piccinini }
\address{Center for Nonlinear Science, University of North Texas,
   P.O. Box 311427, Denton, Texas 76203-1427}
  
\begin{abstract}
We study the spreading of epidemic in a network of individuals who may either contract a disease through sexual contact with the infected nearest neighbors or use safe sex procedures under the influence of neighbors who are already adopting precautions  in their sexual intercourses.  We show that 
both interaction between susceptible $S$ and infected individuals $I$ and the imitation of safe sex procedure, a form of sociological interaction between susceptible $S$ and individuals using safe sex approach $V$, may lead to a phase transition. If the spreading of epidemic is in the supercritical condition, corresponding to an unlimited growth of infection, the interaction between $S$ and $V$ must reach the supercritical condition to generate control of the spreading of infection, and bring the system to criticality. Adopting a theoretical perspective 
similar to the widely used multilayer complex networks, we study the case where the epidemical network is under the influence of a sociological debate on whether to use safe sex or 
not.    We show that at criticality this debate generates clusters of individuals in favor of safe sex techniques and clusters of individuals opposing the use of these procedures. We study the influence of a sociological debate on whether to use safe sex or not, on the spreading of sexually transmitted infections. We show that as a consequence of this debate in the epidemic network a pattern emerges mirroring the structures of the  sociological network. 
Finally we introduce a feedback of the epidemic network on the sociological network and we prove that as a result of this feedback the sociological system undergoes a process of self organization  maintaining it at criticality. We hope that these results may have the important effect 
 of giving interesting suggestions  to the behavioral psychologists and information scientists actively involved in the analysis  of the social debate on the moral issues 
connected to sexual activities.  
\end{abstract}

\maketitle

\section{Introduction}

The transmission of epidemic is a subject of great current interest that can be interpreted as a dynamical process in complex 
networks \cite{book,criticalphenomena}. The spreading of epidemic through a complex network can be used to identify the influential spreaders who are not necessarily the hubs of the complex networks \cite{havlin}. The recent study made by Zoller and Montangero \cite{montangero} with a method of analysis of the world wide web \cite{furini} reveals the inadequacy of the conventional Susceptible Infected Susceptible (SIS) model that has to be modified in such a way as to take into account complex social dynamics. This suggests a connection with the subject, currently in a phase of  rapidly  increasing interest,  of multilayer networks \cite{lastfrontier},  in a form, however, that should imply a perspective different from that suggested by the pioneering paper of Ref. \cite{stanley}. The articles \cite{losalamos1,losalamos2} moving along these challenging directions seem to focus on the topology of the links connecting different layers. Here we follow a direction that is, to some extent, the reverse of the procedure suggested by these articles. The sociological level and the epidemical level of this article are two distinct networks that in the preliminary research work here illustrated are assumed to have the simple topology of two-dimensional regular networks. However, the interaction between the units of this network generates an effective network of strongly correlated units with the structure of a scale-free network. We refer to the earlier work of Ref. \cite{correlation2}  as an example of the directions we want to follow. The authors of this article showed that, although the interacting units are the nodes of a regular two-dimensional network, at criticality a scale-free network of strongly correlated units emerges. In this article we show that the criticality of the sociological level is transmitted to the epidemical level and, as a consequence, on the basis of the results of Ref. \cite{correlation2} we generate two coupled layers of scale-free networks. 

The study of epidemic outbreak, interpreted as phase transition processes, is an interesting branch of epidemiology, discussed   for instance, by Refs. \cite{book,criticalphenomena}. Of notable interest for our discussion is the work of Stollenwerk and Jansen \cite{criticalityinepidemics} and we shall refer to this interesting work to show the connection between our work and the current research work in the field of epidemics, where  criticality is widely studied, mainly from a topological point of view 
with probably deep connections with Self Organized Criticality \cite{soc}.
In this article  we study the influence that sociological criticality 
has on epidemic criticality using the concept 
of temporal complexity \cite{fabio,temporalcomplexity} that we plan to show in action, for the first time, to the best of our knowledge, in the field of epidemics. This study, in a sense,
is a contribution to address the challenge
that the authors of Ref. \cite{perc} set for investigation work
in the field of evolutionary game theory. The novelty of our approach is that although the players interact locally, as in the 
traditional game theory \cite{eshel}, a global behavior favoring the adoption of safe sex techniques, our counterpart of altruism and cooperation, may emerge for reasons totally different from those advocated by the current game theories. In the conventional game theory altruism may survive because altruists tend to clump together in groups, thereby making them survive and expand due to the larger payoff for society \cite{eshel}. This is the essence of the well known work of Nowak and May \cite{nowakmay}.  In this paper we prove with analytical and numerical arguments that in addition to this important property another exists, ignored so far by the researchers in the field of evolutionary games. This is the transmission of the imitation-induced altruistic choices to long distances, due to the long-range correlations associated to criticality.  A selfish unit $B$ may adopt an altruistic choice made by a far away unit $A$ with a delay time much smaller than the time necessary for the information to travel from $A$ to $B$. The systems under study rest on the action of a finite number of units, and, as a consequence, critical slowing down is converted into temporal complexity \cite{temporalcomplexity} and the criticality-induced fluctuations  establish an 
efficient information transfer that should not be confused with an information wave, being based on the criticality-induced locality breakdown \cite{fabio}. 

\section{Two-state model} \label{twostate}

In this section, with the help of a simple two-state model we
illustrate the difference between critical slowing down and temporal complexity. 

\subsection{Critical slowing down}
Let us consider the case where the $N$ units of a network have to make a decision on whether to select the state $A$, which contains the fraction $q$ or the state $B$, which contains the fraction $p = 1-q$ of the entire population. The master equation
for the time evolution of the two probabilities $p$ and $q$ is given by
\begin{equation} \label{first}
\dot p = -\gamma p + \omega q
\end{equation} 
\begin{equation}
\dot q = -\omega p + \gamma q, 
\end{equation}
which are, of course, compatible with the condition $p + q = 1$ at all time. The master equation
is equivalent to the single equation
\begin{equation} \label{tochange}
\dot p = -(\gamma + \omega) p + \omega,
\end{equation}
which is derived from Eq. (\ref{first}) by replacing $q$ with $1-p$. 
In the absence of the natural tendency to imitation that we hypothesize
 to characterize human beings \cite{west},  namely when $\omega = 0$, a unit in the state $A$ will remain there forever, and a unit in the state $B$ will jump to the state $A$ with a finite transition rate $\gamma$.
 In this case all the units end in the state $A$. 
 
 We hypothesize
 that imitation is a typical property of  the individuals of a human society. Consequently we assume that the transition probability of a given unit from the state $A$ to the state $B$ does not vanish, but it is 
 given by
 \begin{equation} \label{noduplicate}
 \omega = K \frac{M_B}{M},
 \end{equation}
 where $M$ is the number of its neighbors and $M_B$ is the number of them in the state $B$. 
 
 In the All-To-All (ATA) case $M$ coincides with $N$.  In the ATA thermodynamic limit,  $N = \infty$, we have
 \begin{equation}
 \omega = K p
 \end{equation}
 and
 the time evolution of $p$ is given, according to Eq. (\ref{tochange})  by
 \begin{equation} \label{thesamestructure}
 \dot p = (K-\gamma) p - K p^2. 
 \end{equation}
 
 With some algebra it is possible to prove that the solution of this equation is:
 
\begin{equation} \label{korosh}
p(t) = \frac{p_0\left(K - \gamma \right)}{\left(K - \gamma - K p_0\right)e^{-\left(K - \gamma \right)t} + K p_0}.
\end{equation} 
Although simple, this equation is a powerful description of the consequences of imitation. 
In fact, it shows that for $K < \gamma$, when imitation is weak, the equilibrium is given by $p = 0$. This indicates that a few individuals in the state $B$ cannot attract a large part of population into this state and these individuals will jump to the state $A$ before attracting any of their neighbors to the state $B$. For $K > \gamma$, on the contrary, the fraction of individuals in the state $B$ is given by
\begin{equation}
p = 1 - \frac{\gamma}{K}. 
\end{equation} 
The individuals who may select for fortuitous reasons the state $B$ may attract many other individuals and for $K\rightarrow \infty$ the whole society ends into the state $B$. 
 
Criticality emerges at
\begin{equation}
K = \gamma.
\end{equation}
In this condition the regression to equilibrium is given by
\begin{equation}
p(t) = \frac{p_0}{(1 + K t)}, 
\end{equation}
which in the long-time limit leads to $p(t) \propto 1/t$.   As a consequence, it takes an infinite time for the system to regress to equilibrium and this phenomenon is well known as critical slowing down.
\subsection{Temporal complexity} \label{twostatetc}

\begin{figure}
 \includegraphics[scale=0.2]{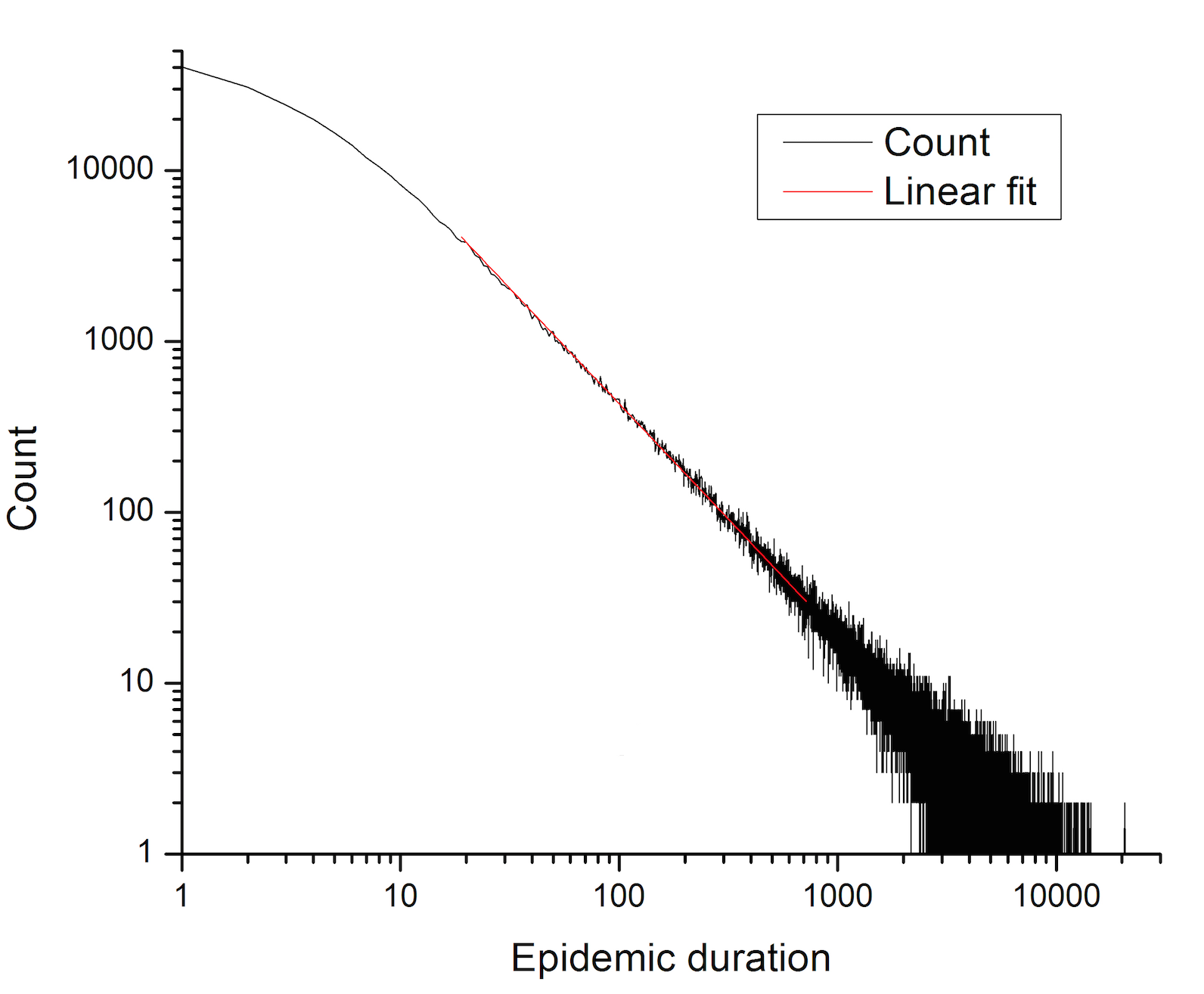}
\caption{\label{swarm2} The distribution density of the time distance between two consecutive epidemic extinctions.  On a two-dimensional lattice with $\gamma=0.05$, $K=0.08$, the initial number of infected was $3$, the total number of trials was $4.2E5$, the threshold was $3$, the slope of the linear fit was $1.35$.}
\end{figure}

In the case of cooperative systems with a finite number of units, critical slowing down is associated to the important property of temporal complexity \cite{temporalcomplexity, mirza}. 
To illustrate temporal complexity let us consider the model of Eq. (\ref{tochange}) with $\omega$ given by Eq. (\ref{noduplicate}) in the crucial case when the number of units is not infinitely large. The algorithm we use to generate the time evolution of the network works with the following prescription.
Let us assume that  a given unit at time $t$ is in the state A. We have to establish whether at the next time $t+1$ it is still in the state A or it jumps to the state B.
 The probability of jumping to the state B is given by $\omega$ of Eq. (\ref{noduplicate}). In the case of a number of units that is not infinitely large $\omega$ reads
 \begin{equation} \label{fluctuation}
 \omega = K (p + f) ,
  \end{equation} 
 where $f$ is a fluctuation of intensity proportional to $1/\sqrt{N}$. 
  In the case of a finite number of units, the process is described by the non-linear Langevin equation
 \begin{equation}
 \frac{d}{dt} p = - (\gamma - K) p - K p^2 + A(p, f),
 \end{equation}
 where $A(p, f)$ is a correction to the merely deterministic description of Eq. (\ref{thesamestructure}) determined by the action of a finite rather than infinite number of units. 
 When $A$ vanishes as a consequence of working with an infinite number of units, the regression to equilibrium is
 characterized by the critical slowing down $1/t$, at criticality, namely, when $\gamma = K$. The work of Ref. \cite{mirza} shows that criticality is a condition favoring fluctuations around the equilibrium value of the mean field, and 
 the distribution density of the time distances between two consecutive origin re-crossings is given by 
  \begin{equation}\label{mu1.5}
 \psi(\tau) \propto \frac{1}{\tau^{1.5}}. 
 \end{equation}
 The two-state model here under study is not symmetric and the regressions to the origin, from either positive or negative values of the mean field become regressions to $p= 0$, with $p$ maintaining the positive value, due to its probabilistic nature. This condition makes the discussion of temporal complexity in this case more complex, thereby opening the possibility that the waiting time distribution density $\psi(\tau)$ may depart from the power index $\mu = 1.5$, as it is shown by Fig. \ref{swarm2} .

\begin{figure}
 \includegraphics[scale=0.2]{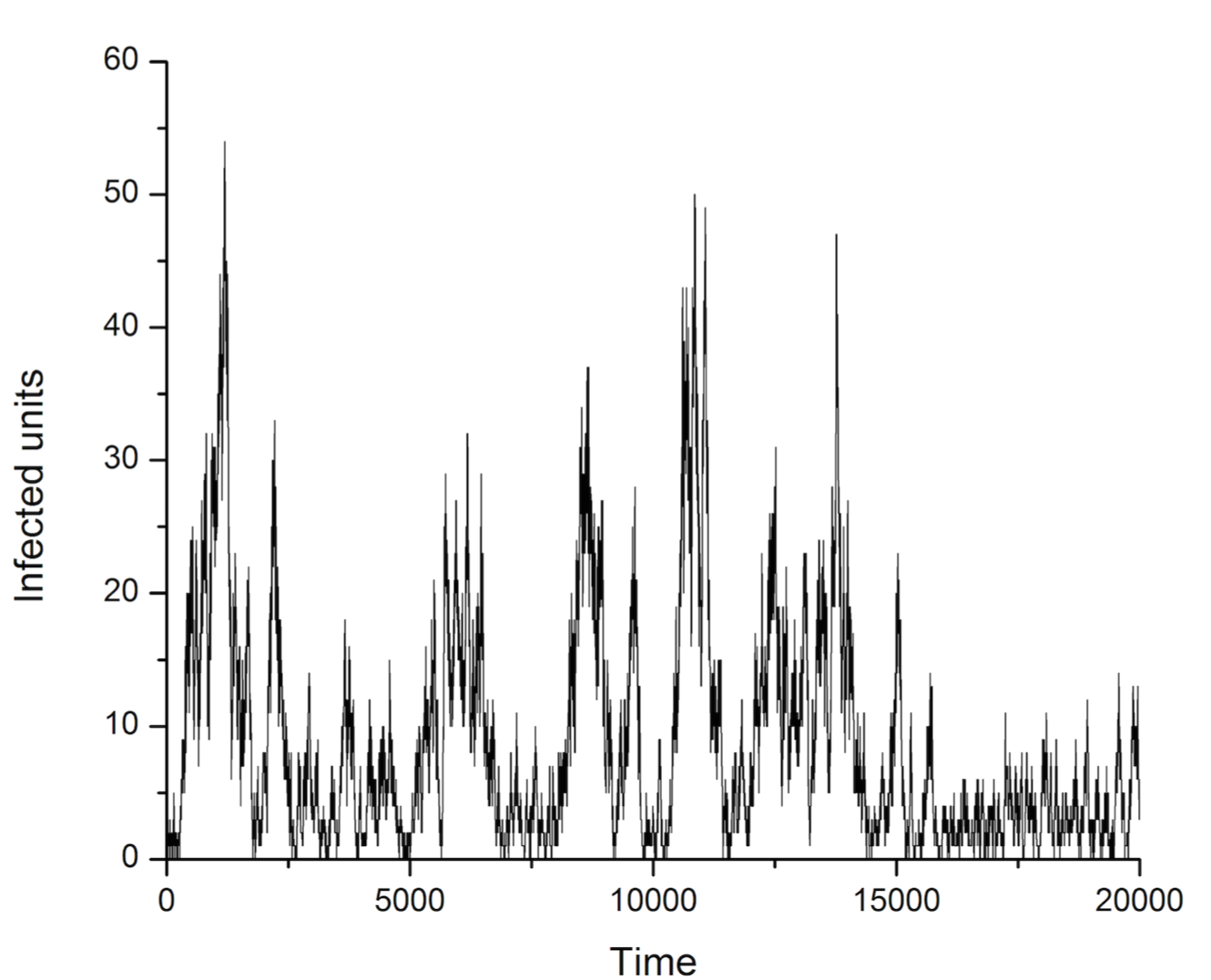}
\caption{\label{swarm1} The infection $I(t)$ as a function of time.  }
\end{figure}

  Note that the state $B$, as we shall see in the next sections,
  may correspond either to  the infection state when we adopt the $IS$ model or to the safe sex state when we adopt the $SV$ model. Here we illustrate the temporal complexity condition using the $IS$ perspective. To establish numerical results compatible with the complex network perspective that we want to promote, we do the numerical calculations using the two-dimensional regular lattice.  In the case of a two-dimensional regular network, where each individual interacts with only four nearest neighbors, the theoretical analysis is more challenging and the power index of temporal complexity departs from the condition discussed in Ref. \cite{mirza}, which yields  $\mu = 1.5$. In Fig.  \ref{swarm1} we see that at criticality infection undergoes frequent collapses. To establish the temporal complexity of these apparently erratic collapses we set a threshold whose crossing corresponds to the extinction of epidemic. In Fig. \ref{swarm2} we illustrate the histogram of the time distances between two consecutive extinctions of epidemic and we find a departure from the ideal 
 condition of free diffusion. 
 It is interesting to notice that
 $\mu = 1.35$ is the same power index as that generated by the organizational collapses of a swarm of cooperative birds, see Ref. \cite{formerfabio}. 
 
Although a theoretical approach to the temporal complexity in the cases of epidemic interest is still missing, there are good reasons to believe that also in this case temporal complexity is the manifestation of a criticality condition, accompanied by the important property of long-range correlation facilitating the communication between distant regions of the same network.   
 
 \section{HIV model} \label{HIV}
 We describe HIV epidemic by means of the following set of equations:
 \begin{equation} \label{firstcapital}
 \dot S = d I + d V - \beta I S - K VS
 \end{equation}
 \begin{equation} \label{secondcapital}
 \dot I = \beta I S - d I
 \end{equation}
 \begin{equation} \label{third capital}
 \dot V = K VS - dV. 
 \end{equation}
 
 Comparing 
 the more complex three-state model of this section to the two-state model of the earlier section, we see that the interaction between the state $S$ and the state $V$ is identical to the interaction between the state $A$ and the state $B$ and we may expect the occurrence of criticality at
 \begin{equation} \label{kd}
 K = d.
 \end{equation}

 However, the criticality of Eq. (\ref{korosh}) may not emerge due to the fact that the individuals of this society in the state 
 $S$ interact also with the individuals of the state $I$, the infected individuals. On the same token, from a mathematical point of view the attraction of $S$ to the state $I$ shares the same structure as the attraction of $S$ to the state $V$. This implies that the parameter $\beta$ plays for the state $I$ the same role as that exerted by the parameter $K$ on the state $V$. In the absence of interaction between $S$ and $V$, 
 the system $SI$ would be characterized by criticality when the 
 condition 
 \begin{equation} \label{bd}
 \beta = d 
  \end{equation}
  applies.    In Section \ref{competingsupercritical} we shall study the competition between the critical condition of Eq. (\ref{kd}) and the critical condition of Eq. (\ref{bd}). Here we  address 
  the case where 
  \begin{equation} \label{b>d}
  \beta > d,
  \end{equation}
  namely, the case when the $SI$ system is in the supercritical regime and a non vanishing  fraction of infected individuals is present in the network.   In the case of no safe sex practices, $K = 0$, the system would generate
  \begin{equation}
  S = \frac{d}{\beta}
  \end{equation}
  and
  \begin{equation} \label{50}
  I = 1 - \frac{d}{\beta}. 
  \end{equation}

 The main question to address is whether the extinction of infection may be a consequence of the attraction exerted on the system by the state $V$. The answer to this question is positive, this equilibrium corresponding to
 \begin{equation} \label{extinction1}
 I = 0
 \end{equation}
 \begin{equation} \label{extinction2}
 S = S^{*} \equiv \frac{d}{K}
 \end{equation}
 
 \begin{equation} \label{extinction3}
 V = V^{*} \equiv 1 - \frac{d}{K}. 
 \end{equation}
 We note that this requires that
 \begin{equation} \label{constraint}
 K \geq d. 
 \end{equation}
 This is a very interesting fact implying that the whole system can be brought back to criticality by the  action of the individuals $V$ when the system $SV$ is in the supercritical condition concerning the adoption of safe sex methods. 
 
 By plugging Eq. (\ref{extinction2}) into Eq. (\ref{secondcapital}), namely adopting a condition very close to equilibrium,  we see that the epidemic infection obeys the equation of motion
 \begin{equation} \label{korfig1}
 \dot I = - R I,
 \end{equation}
 where 
 \begin{equation} \label{korfig2}
 R \equiv d \left(1 - \frac{\beta}{K}\right). 
 \end{equation}
 Thus, we conclude that for $K > \beta$ we have an exponential extinction of $I(t)$ and for $K < \beta$ a complex epidemical dynamics.
 
 By focusing our attention on Eq. (\ref{secondcapital}) we notice that the time derivative of $I$ vanishes 
 when
 \begin{equation} \label{mystery}
 K = \beta .
 \end{equation}
 On the other hand, due to Eq. (\ref{b>d}), Eq. (\ref{mystery}) fits the supercritical condition of Eq. (\ref{constraint}), 
  
  Is the condition of Eq. (\ref{mystery}) associate to a critical slowing down? To answer this important question
  we study the time evolution of $I$, $s \equiv  S - S^{*} $ and $v \equiv V - V^{*}$. It is straightforward to prove
  that Eqs. \ref{firstcapital}, \ref{secondcapital} and \ref{third capital} yield
  
   \begin{equation} \label{firstnocapital}
 \dot s = - \beta I s - \beta s + ds - \beta vs
  \end{equation}
  
  \begin{equation} \label{secondnocapital}
 \dot I = \beta I s \end{equation}

 \begin{equation} \label{thirdnocapital}
 \dot v = \beta s - ds + \beta vs. 
 \end{equation}
 
 Note that, as it must be,  $V + S + I$ remains constant and $v + s + I =0$.
 Using these properties we write Eq. (\ref{firstnocapital}) as
 \begin{equation} \label{newfirst}
 \dot s = - (\beta - d) s + \beta s^2.
 \end{equation}
 Eq. (\ref{newfirst}) has the same structure as Eq. (\ref{thesamestructure}), thereby leading us to
 \begin{equation} \label{finalfors}
 s(t) = \frac{s_0(\beta - d)}{e^{(\beta - d)t}\left(\beta - d - \beta s_0\right) + \beta s_0}.
 \end{equation}
 We are now in the right position to establish the regression to equilibrium of $I(t)$, moving from 
 an out of equilibrium condition $I_0 >0$. In fact, using
 Eq. (\ref{secondnocapital}) 
 we get
 \begin{equation}
 \ln \left(\frac{I(t)}{I_0}\right) = \beta \int_{0}^{t} dt' s(t')
 \end{equation}
 and plugging into it Eq. (\ref{finalfors}), after some algebra we get
 \begin{equation} \label{crucial} 
 I(t) = \frac{I_0(\beta- d) }{\left(\beta - d - \beta s_0\right) + \beta s_0 e^{-\left(\beta - d\right)t}}.
 \end{equation}
 
  This result makes us conclude that the critical slowing down in this case is replaced by an exponential decay from the initial condition $I_0$ to the smaller value  $I_0(\beta-d)/(\beta - d - \beta s_0)$. Note that $s_0< 0$, making $I$ reach
  at infinite time a value smaller that the initial value.  Note also  that if we assume that $I_0$ and $s_0$ are very small, the curve $I(t)$ at criticality remains virtually constant. 
  and \emph{at criticality} $I(t)$ remains virtually constant.
 
  \begin{figure}
 \includegraphics[width=0.4\textwidth,scale=0.5]{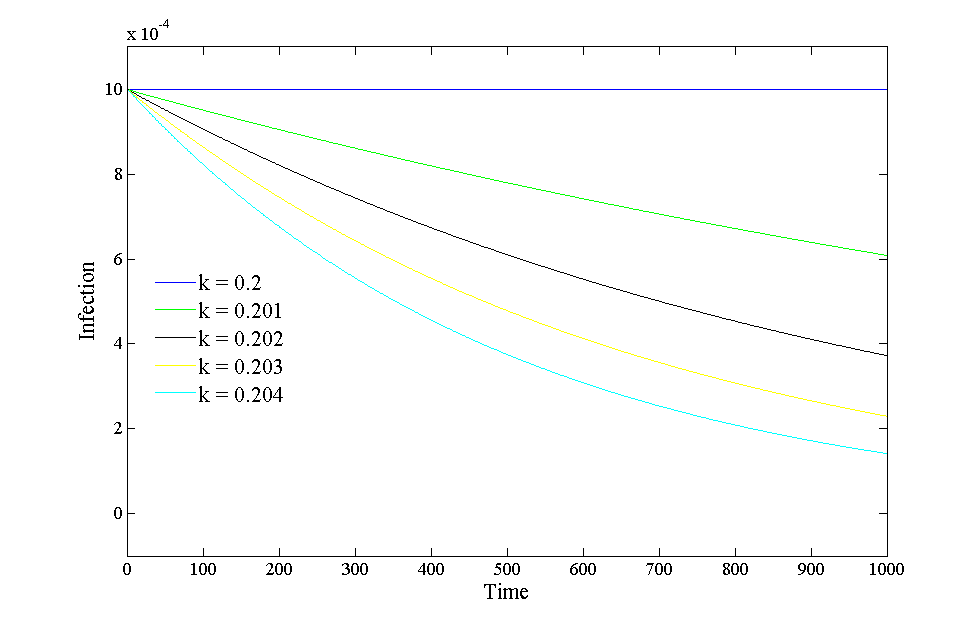}
\caption{\label{fig1} The infection $I(t)$ as a function of time. The curves with $K > \beta$ are exponential functions fitting Eq. (\ref{korfig1}) and Eq. (\ref{korfig2}) with $\beta = 0.2$, $d = 0.1$ and $I_0=0.001$. The curve $K = \beta$ fits the theoretical prediction of Eq. (\ref{crucial}). }
\end{figure}

\section{Two competing super-critical processes} \label{competingsupercritical}

It is interesting to study the case when the both the system $S I$ and the system $SV$ are at criticality, namely when 
both condition (\ref{bd}) and condition (\ref{kd}) apply.  Let us assume that  $I(0)$ is small but positive. To fit the condition
$v + s + I = 0$, both $v$ and $s$ must be negative.  To predict the evolution of $I$ in this condition we study Eq. (\ref{crucial}) for $\beta \rightarrow d$. Setting $x \equiv \beta -d$ and $\lambda = - \beta s_0$ we get 
\begin{equation}
\lim_{x \rightarrow 0} \frac{I_0 x e^{xt}}{e^{xt}\left(x +\lambda \right) - \lambda} = \lim_{x \rightarrow 0} \frac{I_0 \left (e^{xt} + tx e^{xt}\right)}{t e^{xt}\left(x +\lambda \right) + e^{xt}} = \frac{I_0}{1 + \lambda t}. 
\end{equation} 
In conclusion,  in this case we have
\begin{equation}   \label{criticalslowing}
I(t) = \frac{I_0}{1 + \lambda t}.
\end{equation}
In this condition we recover the same critical slowing down as that of the two-state model. It is thus evident that the 
numerical treatment of a  network with a finite number of units will lead to the emergence of temporal complexity, as discussed in detail in Section \ref{twostatetc} and illustrated by Fig. \ref{swarm1} and Fig. \ref{swarm2}. 

 \begin{figure}
 \includegraphics[width=0.4\textwidth,scale=0.5]{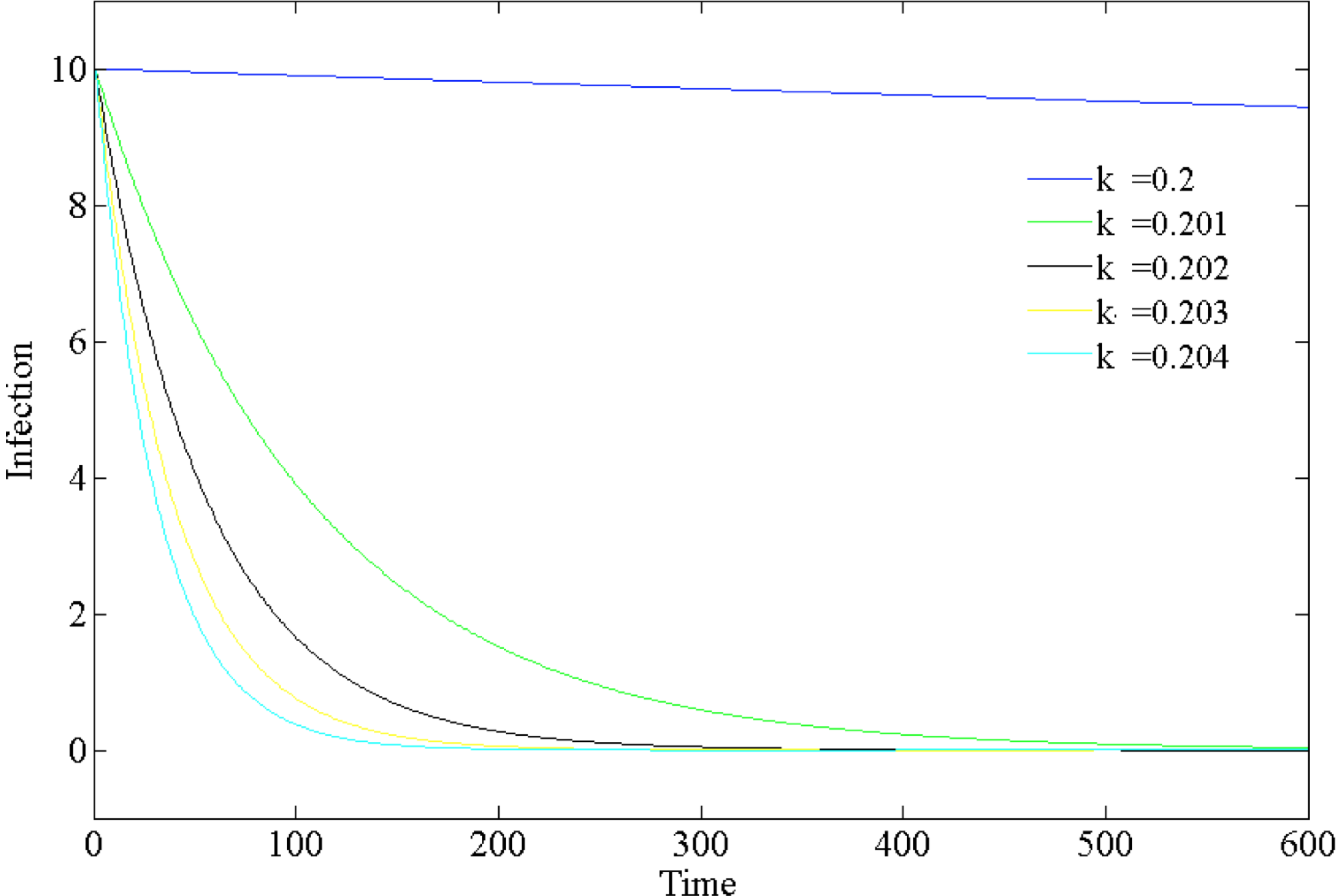}
\caption{\label{fig1} The infection $I(t)$ as a function of time. The curves with $K > \beta$ are exponential functions fitting Eq. (\ref{korfig1}) and Eq. (\ref{korfig2}) with $\beta = d$. The curve $K = \beta$ fits the theoretical prediction of Eq. (\ref{criticalslowing}).  For all curves $\beta=0.2$, $d=0.1$ and $I_0=0.001$.
}
\end{figure}

\section{Decision Making Model}
There is a strong evidence of increasing interest  for the issue of homosexuality and for its connection with sexually induced infections and diseases \cite{HIV,mark}. A detailed discussion of this important issue is beyond the purpose of this article. We limit ourselves to notice that the influence of social network on the epidemic network is a problem of great interest to understand HIV transmission patters and to make intervention to reduce their risk \cite{HIV}. We ignore
the complex psychological process of depression-HIV risk \cite{mark} that would require the addition of further networks in the process that we study. At the very preliminary level of this article we assume that  the sociological network is a two-dimensional regular lattice and that each unit has four nearest neighbors.  We adopt the Decision Making Model (DMM) of ref. \cite{west,temporalcomplexity}. Note that at this level the individuals have to make a choice between two distinct opinions, represented by the states $|1>$ and $|2>$. The state $|1>$ is the pro safe sex state, and the state $|2>$ represents the opposite opinion. The real adoption of safe sex methods or not occurs at the epidemic level and is discussed in Section \ref{DMMdrivingEPIDEMIC}. An individual in the state $|1>$ makes a transition to the state $|2>$ with the
rate
\begin{equation}  \label{prosex}
g_{1 \rightarrow 2} = g_0 exp\left(-K\left(\frac{M_1 - M_2)}{M}\right)\right).
\end{equation}
An individual in the state $|2>$ makes a transition to the state $|1>$ with the rate
\begin{equation} \label{against}
g_{2 \rightarrow 1} = g_0 exp\left(-K\left(\frac{M_2 - M_1)}{M}\right)\right).
\end{equation}
$M$ is the number of neighbors of the decision making unit, $M_1$ is the number of neighbors in the state $|1>$ and $M_2$ is the number of neighbors in the state $|2>$. 
The imitation property is evident. An individual pro-safe sex 
delays the transition to the opposite decision if the majority of its neighbors are making pro-safe sex decision, Eq. (\ref{prosex}), and, on the same token, an individual against adoption of safe sex, delays the transition to the opposite decision  if the majority of its neighbors are making the same decision, Eq. (\ref{against}).  If the majority of neighbors of an individual make the opposite decision, the change of decision occurs earlier than in the absence of imitation interaction with the nearest neighbors. 
Due to the fact that, as earlier mentioned, the network is a regular two-dimensional lattice, we have  $M = 4$. We set also the condition $g_0 = 0.01$ and,  more importantly,   $K=1.7$, which is known to make criticality emerge \cite{west,temporalcomplexity}. 

\begin{figure}
 \includegraphics[scale=0.2]{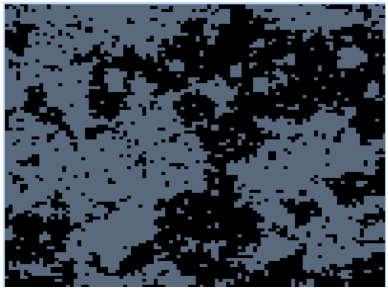}
\caption{\label{debate} The gray regions represent the communities favoring the use of safe sex practices. The black regions correspond to the communities opposing the use of safe sex methods. }
\end{figure}

The interested readers can find additional information in Refs.\cite{west,temporalcomplexity}. However, for their convenience, we show here some of the important properties of DMM that will make easier for them to understand the connection between epidemic and sociological criticality.  First of all,  at criticality, in accordance with 
Refs. \cite{west,temporalcomplexity}, we find the clusters of 
Fig. \ref{debate}. The debate on whether to use a safe sex approach or not at criticality splits the two-dimensional network into clusters, some corresponding to communities favoring and others to communities objecting the safe sex approach.

\begin{figure}
 \includegraphics[width=0.4\textwidth,scale=0.2]{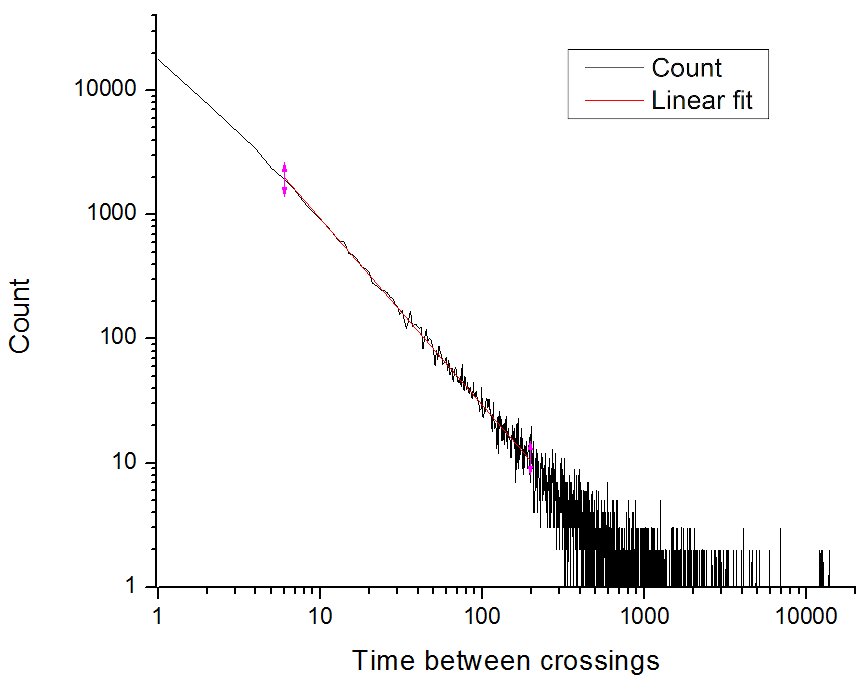}
\caption{\label{freecrit} Distribution density of the time distances between two consecutive origin re-crossings.  The simulation was done on a two-dimensional lattice with $K=1.7$, $g_0=0.01$, a square lattice $100$ nodes on a side, recording $60725$ crossings, in a total of $2.15E7$ timesteps, the linear fit has a slope of $1.50$.}
\end{figure}

The clusters are not static but have complex dynamics that are described in detail in Ref. \cite{west}. 
The imitation process is local, but at criticality long-range effects occur \cite{correlation2}. The long-range correlation has important consequences on the information process, which, using an intuitive metaphor, is equivalent to establish a correlation between an agent moving from the state $|2>$ to the state $|1>$ in Dallas and another agent making the same change of opinion in Los Angeles. As clearly explained in the recent work of Ref. \cite{fabio}, temporal criticality \cite{temporalcomplexity} is the ingredient behind this form of intelligence and for this reason, we devote Fig. \ref{freecrit} to illustrate the temporal complexity corresponding to the clusters of Fig. \ref{debate}. 
For the reader's convenience we show also the numerical results concerning temporal complexity \cite{temporalcomplexity} associate with the criticality condition.  It is interesting to stress that in this case temporal complexity is associate to the power index $\mu = 1.5$ of the distribution density of the  time distances between two consecutive origin re-crossing, a value 
that is strongly supported by the recent theoretical discussion of Ref. \cite{mirza}, whereas the exact value of the corresponding parameter for the two-state model of Section \ref{twostatetc}, as there mentioned,  is still the object of investigation.  We invite the readers to compare Fig. \ref{freecrit} to Fig. \ref{swarm2}, since the main goal of this article is establish a correlation between the sociologic, Fig. \ref{freecrit}, and epidemic criticality, Fig. \ref{swarm2}, 

\section{Epidemic network driven by the sociological network} 
\label{DMMdrivingEPIDEMIC}

In this section we study the influence that the sociological network exerts on the epidemical network. This is done as follows. The $S$ units of the epidemic level corresponding to the state $|1>$ of the DMM level imitate their neighbors in the state $V$ with the imitation parameter $K_{high} = 0.4$, which is well beyond the corresponding criticality value, which is found numerically to be $K = 1.5 \gamma= 0.15$ with $\gamma = 0.1$.  In fact, 
we set $\beta = 0.2$ and $d = 0.1$.  We are in the epidemic supercritical condition of Eq. (\ref{bd}) which, according to the theoretical prediction of Eq. (\ref{mystery})   would require the value $K = 0.2$ for the criticality of the model $SIV$ to occur. 
The imitation strength $K_{high} = 0.4$ would correspond  to a virtual extinction of epidemic, if all the susceptible individual adopted such a high imitation value.  

\begin{figure}
 \includegraphics[scale=0.2]{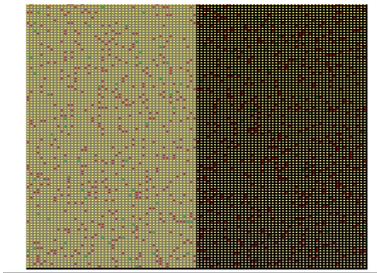}
\caption{\label{spreading1} This figure represents the initial DMM condition. The units on the left half are in favor of safe sex, and the units on the right half are against the adoption of safe sex methods. The red dots correspond to the infected individuals. The green dots are the individual adopting safe sex precautions. The yellow dots are the susceptible individuals. }
\end{figure}

Unfortunately, the $S$ units of the epidemic level corresponding to the sites of the DMM level where the 
no-safe sex decision is adopted, do not set any attention to their nearest neighbors adopting safe sex methods. This is realized by setting for them $K = 0$.  According to the discussion of Section (\ref{HIV}), $K = 0$ would make the system $SIV$ identical to the system $SI$ in the supercritical case, insofar as $\beta = 0.2$ and $d = 0.1$, thereby realizing
the condition of Eq. (\ref{b>d}) which, according to Eq. (\ref{50}) is proven in this article to corresponds to a $50$ percent of infected individuals. As a result of this interaction between the DMM and the epidemic level,  the DMM criticality is transmitted
to the epidemic level, as shown in Fig.  \ref{spreading1}.

For representation convenience the two networks are superimposed the one to the other. In Fig. \ref{spreading1} the left half of the sociological network is a community of individual in favor of the safe sex practices, whereas the individuals of the right half object the use of safe sex methods. The individuals of the DMM system 
are influenced by their nearest neighbors. Thus, the border between the two regions fragments into a more and more complex patterns upon increase of time. The advocates of safe sex methods penetrate the right region and the individual objecting the use of safe sex techniques can convince individuals of the left region to change mind. 
 
\begin{figure}
 \includegraphics[scale=0.2]{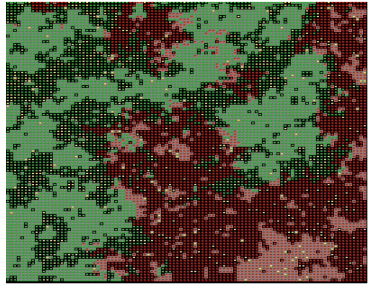}
\caption{\label{spreading2} The red dots correspond to the infected individuals. The green dots are the individuals adopting safe sex precautions. The yellow dots are the susceptible individuals. The red and the green dots look  light 
in the clusters corresponding to the gray regions of Fig. \ref{debate}  and dark green and dark red in the region corresponding to the gray regions of Fig. \ref{debate}}
\end{figure}

 As a result of the debate that we assume to occur at criticality, we expect that  after some time the same clustering structure as that illustrated in Fig. \ref{debate} emerges. This expectation is confirmed by the results illustrated 
 in Fig. \ref{spreading2}.  The details of this figure show additional interesting properties. We see that the number of susceptible individuals is large in the dark green regions. These are the regions where the green clusters generated by 
 the social approval of safe sex methods overlap with the dark clusters of the  individuals opposing the adoption of safe sex precautions. The explanation  of this property is that imitation effect that leads the susceptible individuals  to adopt safe sex procedures  is quenched by social environment of these individuals.  In spite of being surrounded by individuals adopting safe sex procedures, the susceptible individuals do not adopt them because the imitation strength is annihilated by the social opinion opposing the use of safe sex. 
 
\section{Feedback of epidemic on the sociological network}
In this section we illustrate an effect that, to some extent, is related to the recent observation of Ref. \cite{scientificreport}. 
The authors of this interesting paper proved that the information network may have the beneficial effect of quenching the epidemic spreading, because the individuals informed about the 
arrival of an epidemic may have recourse to precautions to reduce the risk of infection. Here we assume that the individuals of the sociological network are aware of the epidemic spreading in their territory and as consequence there may be a bias in  favor of safe sex, which contributes to quench the epidemic spreading as in the case of \cite{scientificreport}. 
, 
The influence of the DMM level on the epidemic level is realized using the same prescriptions as in Section \ref{DMMdrivingEPIDEMIC}. The DMM unit in favor of safe sex
lead the corresponding epidemic units to adopt $K_{high} = 0.6$ and $K_{low} = 0$. Note that $K_{high}$ here larger than 
the value adopted in Section \ref{DMMdrivingEPIDEMIC}, which rests on $K_{high} = 0.4$. 

The feedback is realized according to the following prescription.
A decision making unit in the state $|1>$ makes a transition to the state $|2>$ with the rate
\begin{equation}  \label{prosex2}
g_{1 \rightarrow 2} = g_0 exp\left(-K\left(\frac{(M_1 - M_2) }{M}-  K_F \left(\frac{(M_I - M_V) }{M}\right)\right)\right).
\end{equation}
This corresponds to making the decision process of Eq. (\ref{prosex}) depend on how many of the nearest neighbors are infected. $M_I$ is the number of infected nearest neighbors and $M_V$ is the number of neighbors adopting precaution in their sexual intercourses.  If 
$M_I > M_V$ the decision making unit is encouraged to maintain the safe sex option for a more extends time. On the same token, for a decision making unit in the state $|2>$ we adopt the prescription 
\begin{equation}  \label{nosex2}
g_{2 \rightarrow 1} = g_0 exp\left(-K\left(\frac{(M_2 - M_1) }{M}-  K_F \left(\frac{(M_V - M_I) }{M_{IV}}\right)\right)\right).
\end{equation}
\begin{figure}
\begin{subfigure}{0.4\textwidth}
 \includegraphics[width=\textwidth,scale=0.2]{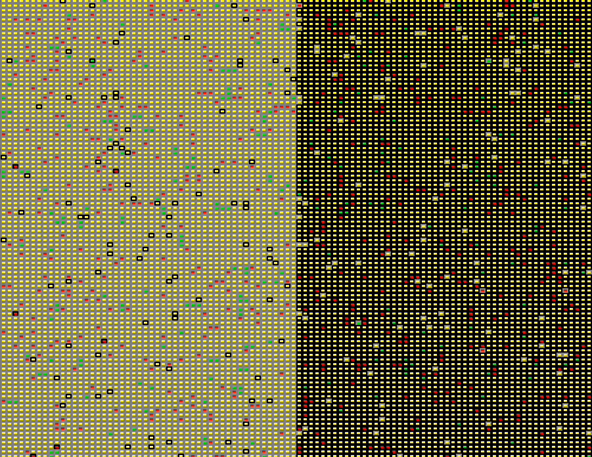}
\caption{}
\label{fig:INITIAL}
\end{subfigure}
\begin{subfigure}{0.4\textwidth}
\includegraphics[width=\textwidth,scale=0.2]{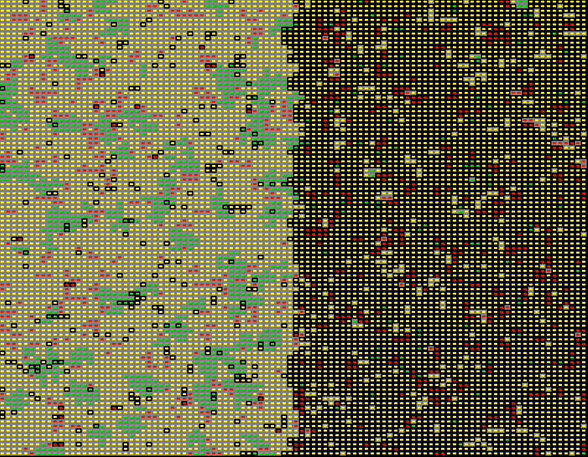}
\caption{}
\label{fig:initial2}
\end{subfigure}
\caption{As in Fig. \ref{debate} the DMM parameters 
are $g = 0.01$, $K = 1.7$. The parameters of the epidemic level are: $d = 0.02$, $\beta = 0.1$. The safe sex imitation strength are $K_{high} = 0.6$, $K_{low} = 0.$ 
The first few steps after the initial condition of Fig. \ref{fig:INITIAL}. }
\end{figure}
This means that a unit in the against safe sex state can make 
a faster transition to safe sex state if $M_I > M_V$. Note that we set $K = 1.7$, which is known to be the DMM critical value \cite{west}. 

To a first sight one may predict that the feedback, favoring the adoption of safe sex will lead to extinction of epidemics. The 
numerical treatment of the process leads us to an even more interesting result. Let us discuss the interesting steps of these complex dynamics. 

Fig. \ref{fig:INITIAL} illustrates the initial condition of the process, virtually identical to that of Fig. \ref{spreading1}.  The left half of the system is in favor of safe sex, while the right side is against safe sex. There are 500 randomly distributed infected units and 200 randomly distributed units that practice safe sex.

Fig. \ref{fig:initial2} shows that, thanks to imitation ($K_{high} = 0.6$ of pro sex units), safe sex practicing units spread in the left part while the infection spreads almost uniformly thanks to the large number of susceptible units.

In the next figure (Fig. \ref{fig:evolution2}), the practice of safe sex is widespread in the left part. We start to notice small ``holes" in the pro safe sex block (left part not completely grey anymore) for units that are not in contact with infection.
We also start to notice nuclei of clusters of pro sex units in the right part, around clusters of infected units. 


In Fig. \ref{fig:widespread}, the left side safe sex is widely practiced, but the majority of units starts to be anti safe sex. Clusters of pro safe sex remain nearby infected areas (this reminds also of current pro/anti vaccine situation \cite{vaccine}).
On the right side disease is spreading thanks to the great number of susceptible units. We notice that there start to be clusters of units that practice safe sex. Clusters of pro safe sex start to develop, too.
\begin{figure}
\begin{subfigure}{0.4\textwidth}
\includegraphics[width=\textwidth,scale=0.2]{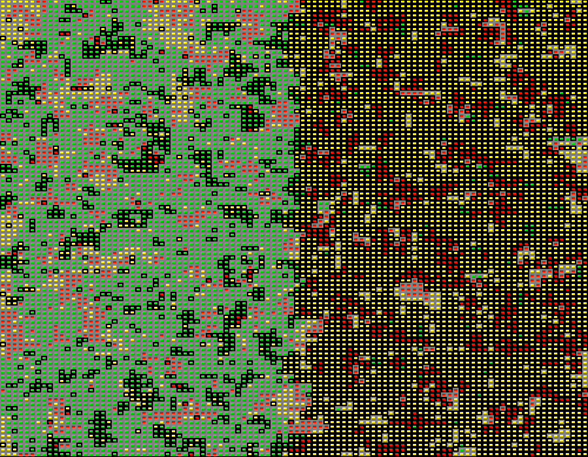}
\caption{}
\label{fig:evolution2}
\end{subfigure}
\begin{subfigure}{0.4\textwidth}
\includegraphics[width=\textwidth,scale=0.2]{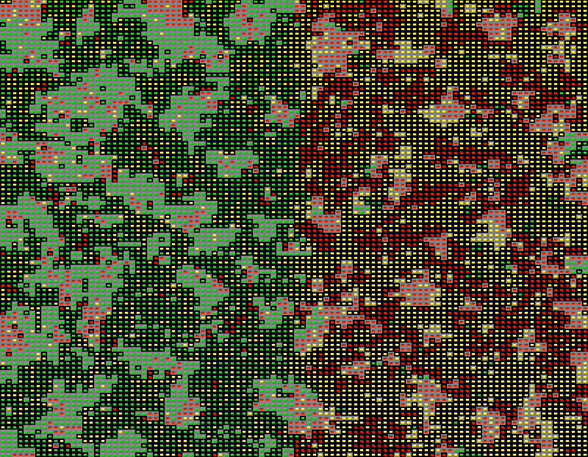}
\caption{}
\label{fig:widespread}
\end{subfigure}
\caption{Emergence of practicing safe sex clusters and pro-safe sex clusters.  
Practice of safe sex is widespread in the left part. }
\end{figure}


This interesting property explains why the conjecture on the feedback-induced extinction of epidemic may not be totally correct. The temporary extinction of epidemic may reduce the adoption of safe sex techniques insofar as it cancels the bias favoring the use of safe sex techniques. As a consequence the movement against safe sex technique may have a significant sociological influence, in the same way as the anti vaccine movement has the effect of reducing the use of vaccination in spite of low cost vaccine being available \cite{vaccine}.

\begin{figure}
\begin{subfigure}{0.4\textwidth}
\includegraphics[width=\textwidth,scale=0.2]{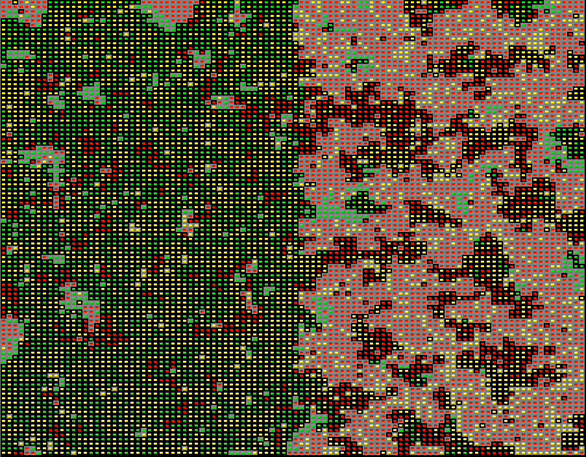}
\caption{}
\label{fig:vaccine}
\end{subfigure}
\begin{subfigure}{0.4\textwidth}
\includegraphics[width=\textwidth,scale=0.2]{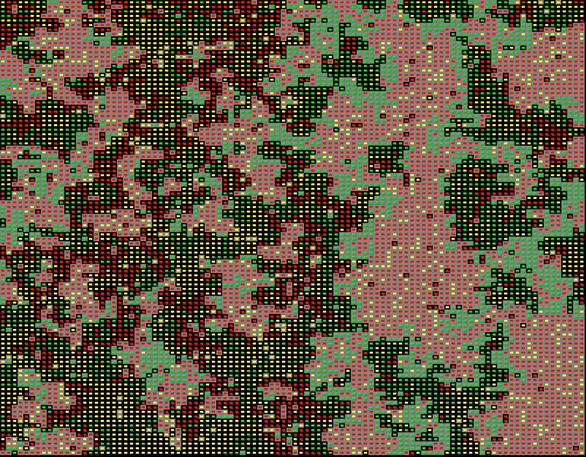}
\caption{}
\label{fig:beforefinal}
\end{subfigure} 
\caption{Infection is spreading in the right side. 
Before the  final}
\end{figure}

Figure \ref{fig:vaccine} shows that because of the spreading of infection in the right side, and thanks to feedback, the pro safe sex units are almost all in that region. Small clusters of safe sex practicing units start to spread.
On the left side pro safe sex units remain only near bigger clusters of infected units. The majority of units is practicing safe sex, but many of them are going back to be susceptible.


In the \ref{fig:beforefinal}, the infection starts to spread again in the left part, and clusters of pro safe sex develop again in the same region in proximity of clusters of infected units.
On the right side the clusters of units that  practice safe sex are continuing their expansion. We notice that the big clusters of infected units are covered by a pro safe sex area: when an infected unit on the border dies, there is a high probability that it will start practicing safe sex as it is in a pro safe sex area and is in contact with units that practice safe sex. This is how clusters of infection decrease their size.


In conclusion, the first sight conjecture that the feedback may make the system depart from criticality due to the extinction of epidemic turns down to not be correct. This is because  the individuals of a territory where epidemic got extinct contribute the debate pro or con safe sex without no bias in favor of sex precautions, and this may have the effect of  preventing the extinction of epidemic, in the same way as the anti-vaccination movement has the effect of preventing the total extinction of infections also in the case where low-cost vaccines are available \cite{vaccine}.

\begin{figure}
\includegraphics[width=0.4\textwidth,scale=0.2]{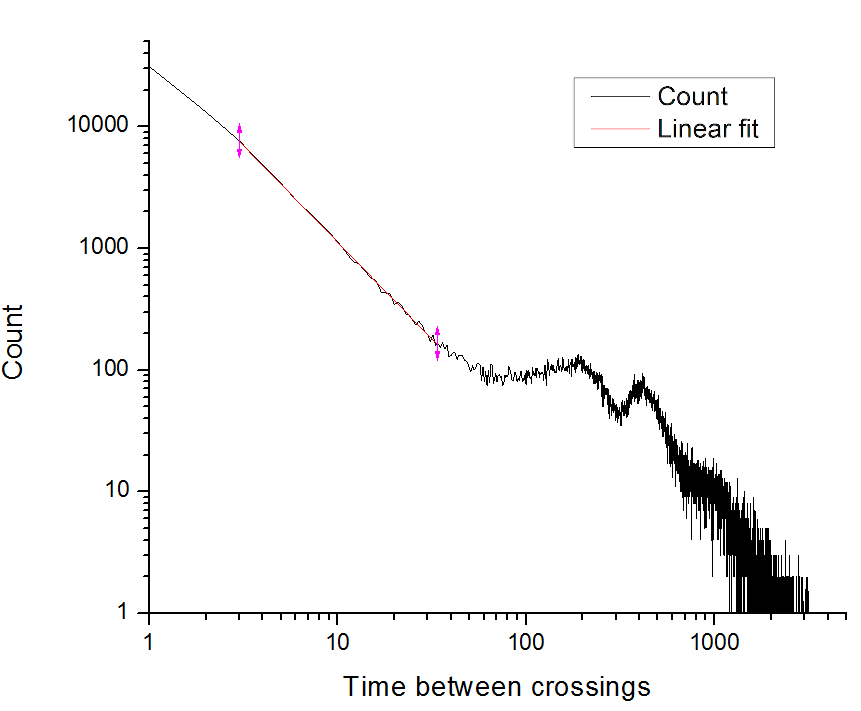}
\caption{\label{feedbackcriticality} Feedback-induced temporal complexity.  The simulation was done on a two-dimensional square lattice with 100 nodes per side, $K=1.7$, $g_0=0.01$, $K_f=1$, $\beta=0.2$, $K_\text{High}=0.4$, $K_\text{Low}=0.0$ and the slope of the linear fit is $1.59$.}
\end{figure}

Does the system remains at criticality? This important question can be answered by following the general direction of the earlier work on DMM \cite{west,temporalcomplexity}. We record
the origin-recrossing of the DMM mean field, in the presence of feedback. The results are illustrated in Fig.  \ref{feedbackcriticality}. Comparing Fig. \ref{feedbackcriticality} to Fig. \ref{freecrit} we note some remarkable properties. The 
extended time region where the distribution density of the time distances between two consecutive regressions to the origin is characterized by the crucial power index $\mu =1.5$ is significantly reduced. The recent research work of \cite{fabio} suggests that this may not totally quench the important properties associated to temporal complexity, long-range correlation and communication efficiency.  In addition to that, we find that the truncation of the power law regime is realized
with a wide shoulder which is not monotonic, with two distinct maxima, suggesting some form of periodic behavior, which is worth of further study. 

\section{Concluding Remarks}

This paper, on one side, rests on the suggestions emerging from the complex patterns generated by the sociological debate
and by the influence of the epidemic feedback on them. A recent example of this way of making new ideas accessible to a wide readership is given, for example, by Ref. \cite{scientificreport2}.  

On the other side, the analytical treatment of Sections II, III and IV, affords a mathematical proof of how a process that is expected  to end up in the supercritical condition can be brought back to criticality by a competing cooperative process in the opposite direction. The imitation of the $IV$ network, which would lead to supercritical condition, in the absence of infected individual, is brought back to criticality by increasing the intensity of the parameter $\beta$ concerning the interaction between $S$ and $I$ individual.

The influence of the feedback of the epidemic level on the sociological level is much more complex, in spite of the fact that 
our model is a extremely simplified if compared to recent observations on the interaction between the epidemic and the sociological level, as discussed for instance in Refs.  \cite{lancet,socialnetwork}, showing that in the case of HIV there is the natural tendency of individual intervention effects to wane over time.  

 However, we are convinced that moving along the directions illustrated in this article  may help the investigators  to 
set on a more solid scientific ground the interdisciplinary research work necessary to establish the influence that society may have on the time evolution of epidemics, especially  HIV epidemic, this being closely connected to the changing sociological perception of what being homosexual and African American homosexual means.  We have in mind the research work of Refs. \cite{HIV} and \cite{socialnetwork} giving strong support to the plausible conjecture that the containment of this form of epidemic requires a wise interaction between the sociological and epidemical network. The DMM sociological network that we use in this paper is a highly simplified version
of the strongly needed more realistic representation that may require the intervention of additional layers that should be used to make this research work fit more properly the current views on the sociological and behavioral perspective behind epidemic spreading.

\emph{ {\bf acknowledgments} We thank Drs. Simone Bianco, Kun Hu and Jaume Kaufman for useful suggestions on the foundation of the epidemic level. We are very grateful to Dr. Mark Vosvick for illuminating directions on the behavioral psychology perspective behind the debate on safe sex and to Professor Victor Prybutok for the statistical analysis of data. }  

.

\end{document}